\documentclass[letter]{aa} 
\usepackage{graphicx}
\usepackage{txfonts}

\newcommand{\beq}{\begin{equation}}
\newcommand{\eeq}{\end{equation}}
\newcommand{\Mdot}{\dot{M}~}

\newcommand{\kms}{\mbox{ km s$^{-1}$}~}

\newcommand{\Mo}{\mbox{M$_{\odot}$}~}
\newcommand{\Moy}{\mbox{M$_{\odot}$ yr$^{-1}$}~}

\begin{document}

\def\etal{et~al.\ }
\def\eg{{\it e.~g.\ }}
\def\ie{{\it i.~e.,\ }}

   \title{Photodissociation in proto-planetary nebulae}

   \subtitle{Hydrodynamical simulations and solutions for low-velocity
    multi-lobes}

   \author{Guillermo Garc\'{\i}a-Segura }

   \institute{Instituto de Astronom\'{\i}a-UNAM, 
Apdo Postal 106, Ensenada 22800,  Baja California, Mexico 
              \email{ggs@astrosen.unam.mx} }

   \date{}

 
  \abstract
   {}
   {We explore the effects of photodissociation at the stages of
   post-asymptotic giant branch stars to find a mechanism able to
   produce multi-polar shapes.  }
   {We perform two-dimensional gasdynamical simulations to model 
the effects of photodissociation in proto-planetary nebulae. }
   {We find that post-asymptotic giant branch stars with 
$\sim 7,000$  K or hotter are able to photodissociate a large amount
of the circumstellar gas.  We compute several solutions for nebulae with
low-velocity  multi-lobes.
We find that the early expansion of a dissociation front is crucial
to understand the number of lobes in proto-planetary nebulae.  }
   {A dynamical instability appears when cooling is included in the
swept-up molecular shell.
This instability is similar to the one found in photoionization fronts, 
and it is associated with the thin-shell Vishniac instability. 
The dissociation front exacerbates the growth of the 
thin-shell instability, creating a fast fragmentation in shells expanding
into media with power-law density distributions such as $r^{-2}$.  }


   \maketitle
%

\section{Introduction}

Several surveys of proto-planetary nebulae (PPNs) and planetary nebulae (PNs)
(\cite{sahai07, chu87, balick87, schwarz92, stanghellini93, 
manchado96a,manchado96b}) show a rich variety of shapes, and they have been
cataloged in a series of morphological classes: bipolar, elliptical, 
point-symmetric, irregular, spherical,  quadrupolar, and multi-polar.
While the first five former classes have been reproduced with 
different scenarios by hydrodynamical and magnetohydrodynamical simulations
(\cite{icke88, icke89, soker89, mellema91, icke92, frank94, dwarkadas96,
rozyczka96,ggs97, ggs99, rijkhorst05, ggs05}), 
the last two former classes have eluded any attempt of modeling,
except the one by \cite{rijkhorst05} and \cite{matt06}. 
In general, all previous scenarios are based on the collimation of winds.
Two examples of low-expanding, multi-polar PPNs  are IRAS 19024+0044 
(\cite{sahai07b}) and  IRAS 19475+3119  (\cite{sahai07c}). It is interesting 
that these two nebulae neither show any kinematic fingerprint in 
their optical spectra nor a perfect symmetry.

\begin{figure*}
\centering
\includegraphics[width=15cm]{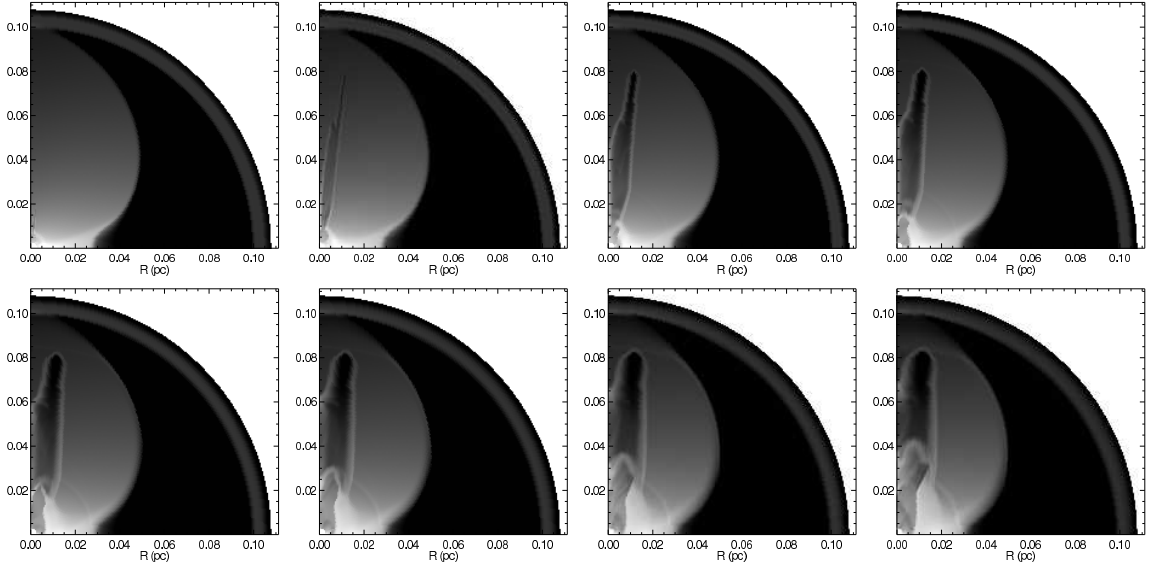}
\caption{Snapshots of gas density at  1, 2, 3, 4 (top), and 5, 6, 7,
8 $\times 10^3$ yr (bottom) (log(min)=-22, log(max)=-18). }
\end{figure*}

Only recently, the hypothesis that
bipolar and multipolar lobes of PNs are the result of ionization    
and illumination was proposed by \cite{kwok10}.
He suggested that multipolar nebulae with similar 
lobe sizes are not caused by simultaneous ejection of matter in 
several directions, but by the leakage of UV photons into those directions. 
Therefore, the  explanation to the multipolar phenomenon does not lie in the
dynamical ejection, but in how such holes were formed. 
The similarity in sizes of the multiple lobes was therefore naturally 
explained, and no simultaneous ejection was required.

According to Kwok's idea, it is necessary to find a mechanism that is able to 
create holes in the surrounding medium of PPNs, different to stellar winds or
collimated outflows of any kind, such as jets originated by accretion disks
(\cite{livio97}),
or the effects from photoionization (\cite{ggs96}), because
post-asymptotic giant branch stars (post-AGBs) are still  not 
hot enough to produce a considerable ionization of the medium.

\cite{diaz98} made an important contribution  to the study of 
photodissociation regions (PDRs) around
main-sequence stars. They outlined the importance of PDRs for 
low-temperature stars (down to 7,500 K), 
because the number of non-ionizing photons in the range 
of  912 \AA $  < \lambda <  $ 1100 \AA \,  is several orders of magnitude 
larger than the ionizing photons ( $\lambda < $ 912 \AA), which are able to 
dissociate the outlying molecular gas, forming  PDRs.

A number of previous studies (\cite{latter00} and references therein) have
outlined the importance of PDRs in the context of PNs. 
However, previous models (\cite{natta98}) have focused on stellar temperatures 
above 30,000  K , or even higher,  such as the case of the Helix Nebula
(\cite{will07}).

Inspired by Kwok's idea, we explore the effects of
photodissociation at the stages of post-AGBs, when stars still have low
temperatures ($\sim$ 7,000 K), i.e., they are still in the so-called
transition time ($5,000 \leq 10,000 $ K), and it is found that 
PDRs are able to explain several
observational features found in PPNs.
The transition time of post-AGBs is not yet fully understood
(\cite{stanghellini00}), since
the mass-lost rate at these stages is basically unknown .
Stellar evolution calculations of PN central stars usually avoid this stage
and start at 10,000 K (\cite{vassi94}), although there is a model 
for $3_{\rm ZAMS} \Mo$ that starts at $\sim$ 6,000 K (\cite{blocker90}).
In the context of the present paper, long transition times $\geq$ 10,000 yr 
(low mass stars) will produce dynamically, developed PDRs, while
transition times shorter than  1,000 yr (massive stars)  do not give
enough time to develop a structured PDR and will form rapidly 
photoionized nebulae.

In this paper, we use the study by \cite{diaz98} as a starting point, 
and merely replace the constant density condition by a numerical integration 
for any density distribution.
The paper is structured as follows: 
Sect. 2 describes the numerical models and physical approaches used in this
study.  
Sect. 3 presents  two-dimensional results of PPNs.  
Sect. 4 discusses the results and gives a summary.

\section{Numerical models and physical approaches}

The simulations have been performed with the magnetohydrodynamical code
ZEUS-3D version 3.4 (\cite{clarke96}), 
developed by M. L. Norman and the Laboratory for Computational Astrophysics. 
This is a finite-difference, fully explicit
Eulerian code descended from the code described in \cite{stone92}.
ZEUS-3D does not include radiation transfer, but we have
implemented a simple approximation (\cite{diaz98}) to derive the 
location of the dissociation front for arbitrary density 
distributions, similar
to the approximation used to locate the ionization front 
(\cite{tenorio79, bodenheimer79, franco89, ggs96}). 
This is done by assuming that dissociation equilibrium
holds at all times, and that the gas is fully dissociated  inside the 
PDR.
One can define the PDR equilibrium radius similarly to the
Str\"omgren radius (\cite{diaz98}),  and the position
of the dissociation  front in any given direction $(\theta ,\phi )$ 
from the photodissociating  source is given by 
\beq
4 \pi \int^{R_{\rm PDR}}_{R_{\rm HII}}   n_{\rm H^0} \, n_{tot} 
\, \alpha_f \, r^2 \, dr \,  = \,  <p> \, S_{\rm D} ,
\eeq
where
$n_{\rm H^0}$ is the density of neutral hydrogen,
$n_{tot} = n_{\rm H^0} + 2 n_{\rm H_2} $ is the total proton density,
$\alpha_f = 1.37 \times 10^{-17} $ (\cite{hollenbach71}), 
$ <p> = 0.15 $ (\cite{abgrall92, draine96}), and
$S_{\rm D}$ the rate of dissociating photons.
In the context of this paper, we assume that $R_{\rm HII}=0$. 
The left side of Eq. (1) is the number of ${\rm H_2}$ molecules 
formed in a given volume, while the right side is the total number 
of photodissociations per unit time, where $<p>$  is the average dissociation
probability for the wavelength range 912-1100 \AA. 

The stellar parameters are taken from a model for $M_{\rm ZAMS} = 2.5 \Mo $
(\cite{vassi94}), and using a black-body approximation similar to
\cite{villaver02} for $T_{\rm eff} = 6-7-8-10 \times 10^3  $ K, 
we obtain  $ log(S_{\rm D}) \sim 42, 43, 44, 45 s^{-1}$.

The models include the cooling curves 
by \cite{dalgarno72} and \cite{macdonald81}.
The cooling cutoff temperature of the molecular gas is set to 10 K. 
Finally, the photodissociated  gas is always kept at $10^3$ K , so no 
cooling curve is applied to the PDRs (unless there is a shock 
inside the PDR). 

Equatorial density enhancements are introduced in the two-dimensional
 models using
Eq. 2 to 5 in \cite{ggs99}, based on the wind-compression 
equations by \cite{bjorkman93}.
Here , $\Omega = v_{\rm rot}/v_{\rm crit}$  gives the proximity of the star
to critical rotation (see Eq. 6 in \cite{ggs99}).
We assume that critical rotation (or close) is achieved at some point
in the blueward excursion (\cite{langer98, heger98}) during the
transition time, similar to the model computed for 12 \Mo (\cite{chita08}).  
Common envelope scenarios probably produce similar effects, although
there are not yet any available  equations in the literature.

\section{Two-dimensional results}

The first two-dimensional model consists of three stages. In the initial stage, 
we initialized the computational domain with a spherical AGB wind with 
$\Mdot = 1 \times 10^{-5} \Moy $ and $v_{\infty} = 10 \kms$. 
Our grid consists of $200 \times 180$ equidistant zones in $r$ and $\theta$,
respectively. The innermost radial zone lies 
at $r_{\rm i}= 2.725 \times 10^{-3} $ pc , and the outermost zone 
at $r_{\rm o} = 0.109 $ pc .  The angular extent is $90^{\circ}$.

In the second stage, we introduce a second wind at the inner boundary  with the 
same parameters, but it is assumed that the star is near critical rotation with 
$\Omega = 0.98$. This is done for a period of $10^4$ yr. 
We do not include here a smooth transition 
(from $\Omega = 0 \longrightarrow 0.98$) for simplicity. 
These two stages are then used as initial conditions for 
a third stage. We set all velocities in the grid to zero. 
This is done to avoid the expansion of the AGB wind out of the grid and
to compute only the expansion of the PDR.

The first snapshot in Fig. 1 (at 1,000 yr) shows  the structure generated by 
both AGB winds, an outer spherical wind almost at the edge of the grid, 
followed by a bipolar wind. 
This bipolarity is owing to the smaller values of the radial 
velocity toward the equator in the AGB wind with critical rotation 
(see Eq. 2 in \cite{ggs99}), which produces an apparent empty space
because of a rarefaction wave in the wind (dark areas in the plot). 
The dark areas
would be filled with gas using a transition, evolving wind with
$\Omega = 0 \longrightarrow 0.98$.

In the third stage, we simulate the conditions of a post-AGB wind near 
critical rotation ($\Omega = 0.98$), decreasing the mass-loss rate down
to $\Mdot = 1.73 \times 10^{-7} \Moy$. Here, we assume that the star has 
reached 7,000 K , and the corresponding UV photons reach 
$S_{\rm D} =  10^{43} $ s$^{-1}$. As before, we artificially set
the wind velocity to zero, just to compute only the contribution from the 
thermal expansion of the photodissociated gas. By doing this, 
we are also assuming at the same 
time that the post-AGB star is not able to drive an 
efficient wind, because the stellar temperature is not high enough to 
produce a line-driven wind.

Figure 1 shows the evolution of the gas density during  8,000 yr. 
During the first thousand years, the lower density
at the polar region allows the PDR to expand in the form of beams, 
digging low-density tunnels, preferentially at the polar axis 
(first snapshot in Fig. 1). The gas is evacuated and pushed 
sideways in those beams because of the sudden increase of the temperature 
(from 10 to 1,000 K) 
in the PDR.  During the second thousand years (second snapshot in Fig. 1), 
the growth of the beam at the polar axis ends because of the lateral 
expansion of the two adjacent beams, blocking the UV field owing to the 
increase of density at the axis. At the end of this phase at 2,000 yr, 
two thin lobes (0.08 pc long) are formed at each hemisphere 
(second snapshot in Fig. 1), producing a quadrupolar shape. 

\begin{figure}
\centering
\includegraphics[width=7cm]{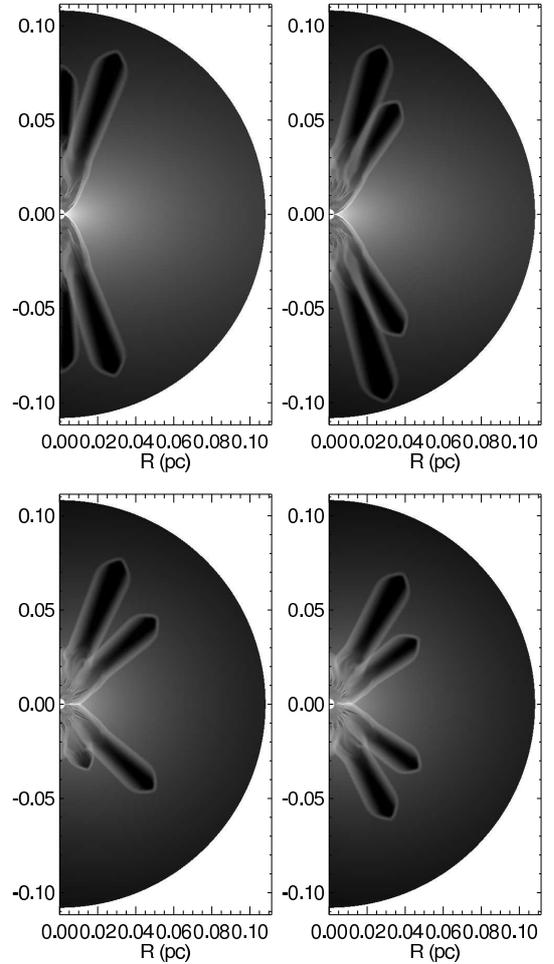}
\caption{ From left to right, top to bottom: gas density of models
80-a, 80-b, 50-a, 50-b at 9,000 yr
of the evolution (log(min)=-22, log(max)=-18).
Model numbers relate the approach to critical rotation (in percent).}
\end{figure}

\begin{table}
\caption{Sample of models}
\label{table:1}
\begin{tabular}{l c c}
\hline\hline
model & $\Mdot_{\rm AGB}$ ($\Moy$)  & $\Omega$ \\
\hline
80-a     &     $0.60 \times 10^{-5}$  &      0.80 \\
80-b     &     $0.50 \times 10^{-5}$  &      0.80 \\
50-a     &     $0.40 \times 10^{-5}$  &      0.50 \\
50-b     &     $0.38 \times 10^{-5}$  &      0.50 \\
\hline
\end{tabular}
\end{table}

Meanwhile, the density gradient between pole and equator in the PDR close 
to the central star produces a  flow directed toward the polar axis. 
This bipolar flow converges at high latitudes, and is able to increase the 
density up to the point of producing a self-blocking of the UV field  
between 2,000 and 3,000 yr.  At this moment, the PDR contracts and 
becomes confined by this flow at $\sim 0.01$ pc,  and the external lobes 
cool down to form hydrogen molecules.  After 3,000 yr, the PDR again expands 
in the polar direction in a bipolar or quadrupolar form 
(the dissociation front has funnel shape). 

The last snapshot in Fig. 1 (at 8,000 yr)  shows the final development
of our model. At 0.02 pc, a high-density blob produced by the convergence of
two oblique shocks defines the position of the dissociation front at the 
polar axis.
Above this blob, the remnant lobes collimate the expansion of the
molecular flow beyond the dissociation front. At 0.045 pc, the molecular 
outflow reaches values of 8.7 \kms.  

The initial expansion of a dissociation front could be very sensitive to the 
conditions found at the density ramp. To illustrate this behavior, we
have computed a sample of  four models in which we have changed the
equator to pole density ratio by changing the stellar rotation
($\Omega = $ 0.80, 0.50 ) , with a
slightly different mass-loss rate for those with the same rotation (Table 1). 
The other computational inputs have the same properties as the former model, 
except that we skip the first stage (the initial spherical wind), 
and directly initialize the computational mesh with the wind near critical 
rotation.  The post-AGB wind is not included either, only the UV field.
The results are shown in Fig. 2, with snapshots at 9,000 yr. 
The angular extent is now $180^{\circ}$ in $\theta$, 
with 360 zones. We have introduced a 1\% random noise in the
initial density to allow for differences between the northern and the southern 
hemispheres, to emphasize that the lobes could reach different sizes and
be formed at different angles, i.e., they are not necessarily symmetric.
The expansion velocity of the lobes in Fig. 2 is lower than 1 \kms
with respect to the AGB wind ($=0$ in the simulations). 

We cannot discuss each model in detail here, but give a short 
summary. We find that the early expansion of a dissociation front is crucial to
understand the number of lobes in the models, and we also find
that the number of lobes is inversely proportional to the equator/pole 
density ratio. A  PDR is able to expand in more directions when the
ratio is low. 
For example, in the extreme case of Fig. 1, the PDR only occupies a 
few cells in the polar region, and the dissociation front digs 
a tunnel only through the polar axis (small solid angle), 
while in the case of model 50-b (Fig. 2), the
initial PDR occupies many cells down to much lower latitudes, and the PDR
expands into a larger solid angle.  In the ideal case of 
a spherical wind (not shown in the figure), the two-dimensional result has
the shape of a daisy flower, because the PDR expands in all directions.
Note that the development of instabilities in a
dissociation front  is crucial to understand the expansion of PDRs 
(see Sect. 4).

\section{Discussion and summary}

\cite{ggs96} noted that a D-type dissociation front preceded by
a radiative shock should be unstable, similarly to the ionization-shock (I-S) 
front instability of a  D-type ionization front. 
By analogy to the I-S front instability (\cite{giuliani79}), 
we called it here dissociation-shock (D-S) front instability. 
The dynamical behavior of the D-S front instability is equivalent to
the one of the I-S front instability (\cite{ggs96, whalen08}), the only 
difference is that the temperature of a PDR  is around one order of magnitude
lower than that of a HII region. This causes a lower pressure
gradient between the molecular and the dissociated gas, and consequently, the
growth, development, and dynamics of the D-S front instability would
be less violent than that of the I-S front instability.

At the onset of the D-S front instability, the CSM  gas is forced to 
cross oblique shocks, and the gas acquires a
transverse component in velocity (\cite{vishniac83}, \cite{maclow93}). 
Thus, the gas piles up in the ``valleys'' (by analogy), enhancing
the column density, but the gas is cleared out from the ``peaks'',
 therefore lowering
the column density. The radiation field notices the redistribution of the gas
and a number of photons can now reach deeper regions in the directions where
the ``peaks'' are located. 
Because the inertia of the swept-up shell decreases
at the ``peaks'', the dissociated gas expands and accelerates the leading shock,
steepening the angle of the oblique shocks. If the CSM has a decreasing slope
in density (such as $r^{-2}$ in AGB winds), 
the growth of this instability becomes catastrophic. 
Dissociation fronts should be also subject to a shadowing instability 
similar to ionization fronts (\cite{williams99, whalen08}), which probably
triggers the development of the D-S front instability. 

We have shown here that photodissociation produces an important 
fragmentation in PPNs via the D-S front instability, and is able to shape 
complicated structures, like quadrupolar and multipolar lobes.  
Then, according to Kwok's idea, photodissociation is the mechanism able to
create holes in the surrounding medium of post-AGBs. 
Once the holes are formed, the post-AGB wind could be channeled through 
them, forming multipolar bubbles. 

In a future article, we will perform calculations using new evolutionary 
tracks for post-AGBs, including other inputs such as stellar winds, 
with resolution studies for the D-S front instability, and with 
3-D solutions.

\begin{acknowledgements}

G.G.-S. thanks the referee, Vincent Icke,  for his valuable comments which 
improved the article, and  thanks Michael L. Norman and the 
Laboratory for Computational 
Astrophysics (LCA ) for the use of ZEUS-3D.  This work has been partially 
supported by the grant DGAPA-UNAM IN100410-3.

\end{acknowledgements}

\end{document}